\documentclass[10pt,aps,prd,twocolumn,nofootinbib,showpacs,showkeys,superscriptaddress]{revtex4-2}

\usepackage{psfrag} \usepackage{mathrsfs} \usepackage{amssymb, bm} 
\usepackage{amsmath, amsthm} \usepackage{epstopdf} 
\usepackage[breaklinks=true]{hyperref} \usepackage{enumerate} 
\usepackage{longtable} \usepackage{subfigure} 
\usepackage{mathrsfs} \usepackage{graphicx} 
\usepackage{natbib,url,textcase}  \usepackage{xurl} 
\usepackage[dvipsnames]{xcolor}
\usepackage[title]{appendix}

\usepackage[english]{babel} \usepackage[T1]{fontenc} \usepackage{comment} 
\numberwithin{equation}{section}

\usepackage{geometry}
\allowdisplaybreaks

\hypersetup{
}

\usepackage{tikz}

\newcommand{\be}{\begin{equation}} 
\newcommand{\ee}{\end{equation}} 

\definecolor{purple}{rgb}{1,0,1} \definecolor{lime}{HTML}{a6CE39} % needs xcolor \newcommand{\lime}[1]{{\color{lime} #1}} 

\newcommand{\orcidicon}{%
	\begin{tikzpicture}
		\draw[lime, fill=lime] (0,0) circle [radius=0.15] 
		node[white] {{\fontfamily{qag}\selectfont \tiny ID}}; 
		\draw[white, fill=white] (-0.0625,0.095) circle 
		[radius=0.007];
	\end{tikzpicture} \hspace{-2mm} }

\newcommand\orcidValerio{{\href{https://orcid.org/0000-0002-2601-1870}{\orcidicon}}}
\newcommand\orcidLuca{{\href{https://orcid.org/0009-0006-3167-4990}{\orcidicon}}}
\newcommand\orcidSantiago{{\href{https://orcid.org/0009-0001-8474-145X}{\orcidicon}}}
\newcommand\orcidAlain{{\href{https://orcid.org/0009-0006-2244-0084}{\orcidicon}}}

    \def\T{{\cal T}} 
\def\K{{\cal K}}
\def \d {{\mathrm{d}}}

\begin{document} \def\theequation{\arabic{section}.\arabic{equation}}

\title{New interpretation of the Minkowski limit of $R^2$ gravity}

\author{Valerio Faraoni\orcidValerio} \email[]{vfaraoni@ubishops.ca} 
\affiliation{Department of Physics \& Astronomy, Bishop's University, 2600 
College Street, Sherbrooke, Qu\'ebec, Canada J1M~1Z7}

\author{Luca Gallerani\orcidLuca} \email[]{luca.gallerani6@unibo.it}
\affiliation{Department of Mathematics, University of Bologna, Piazza di 
Porta San Donato 5, 40126 Bologna, Italy}

\author{Alain Maltais-Gosselin\orcidAlain} \email[]{amaltais26@ubishops.ca} 
\affiliation{Department of Physics \& Astronomy, Bishop's University, 2600 
College Street, Sherbrooke, Qu\'ebec, Canada J1M~1Z7}

\author{Santiago Novoa-Cattivelli\orcidSantiago} 
\email[]{snovoa23@@ubishops.ca} 
\affiliation{Department of Physics \& Astronomy, Bishop's University, 2600 
College Street, Sherbrooke, Qu\'ebec, Canada J1M~1Z7}

\author{Uri Gorman} \email[]{ugorman24@ubishops.ca}
\affiliation{Department of Physics \& Astronomy, Bishop's University, 2600 
College Street, Sherbrooke, Qu\'ebec, Canada J1M~1Z7}

\begin{abstract}

It is well-established that the Minkowski limit of pure $f(R)=R^2$ gravity 
breaks down, unlike that of full Starobinsky theory $f(R)=R+\alpha R^2$. 
We provide a novel interpretation of this phenomenon using the recent 
thermal analogy between scalar-tensor gravity and Eckart's relativistic 
dissipative fluids. In this framework, we show that approaching the 
Minkowski background corresponds to a diverging effective ``gravitational 
temperature''  along certain non-stationary FLRW 
trajectories, while the gravitational temperature remains identically zero 
along the exact de Sitter family. This perspective naturally rephrases 
the strong coupling problem as a thermal singularity, demonstrating that 
$R^2$ gravity departs infinitely far from General Relativity rather than 
recovering it.

\end{abstract}
\maketitle

\section{Introduction} 
\label{sec:1}
\setcounter{equation}{0}

There are many reasons to believe that gravity deviates from General 
Relativity (GR) at high energies, and quite possibly also in the infrared. 
Indeed, the acceleration of the cosmic expansion at the present epoch, 
discovered using Type Ia supernovae, has led to introducing dark energy 
(see \cite{Amendola:2015ksp} for a review of models) which, however, is 
completely {\em ad hoc}. It is more appealing to explain the present 
cosmic acceleration by modifying gravity at large (cosmological) scales. 
Much effort has 
gone into explaining away dark energy with $f(R)$ gravity, a class of 
scalar-tensor theories in which the 
Einstein-Hilbert gravitational Lagrangian density $R$ (the Ricci scalar of 
spacetime) is generalized to a non-linear function $f(R)$ (see 
\cite{Sotiriou:2008rp,DeFelice:2010aj, Nojiri:2010wj} for reviews).

Regardless of the infrared limit, modifications of GR are needed in the 
ultraviolet regime, where quadratic quantum corrections are the first to 
appear  \cite{Stelle:1976gc, Starobinsky:1980te, Starobinsky:1979ty}. Pure 
$R^2$ gravity is the subject of a fairly large literature. First, it is 
believed to be renormalizable and asymptotically free \cite{Stelle:1976gc, 
Starobinsky:1980te, Starobinsky:1979ty, Fradkin:1981hx, Adler:1982ri, 
Avramidi:1985ki, Rinaldi:2014gua}. Second, it is a special case of more 
general (metric) $f(R)$ theories of gravity described by the 
action\footnote{We adopt the notation of Ref.~\cite{Wald:1984rg}: the 
metric signature is ${-}{+}{+}{+}$ and we use units in which the speed of 
light $c$ and Newton's constant $G$ are unity, $g$ is the determinant of 
the spacetime metric $g_{ab}$, which has covariant derivative 
$\nabla_a $, and $\Box \equiv g^{ab} \nabla_a\nabla_b $.}
 \be
S=\frac{1}{16\pi} \int d^4 x \, \sqrt{-g} \, f(R) +S^\mathrm{(m)} 
\ee
(where $S^\mathrm{(m)} = \int d^4x \, \sqrt{-g} \, {\cal 
L}^\mathrm{(m)}$ is the matter action), which are extremely 
popular to describe the present acceleration of the cosmic expansion 
without invoking dark energy (see \cite{Sotiriou:2008rp,DeFelice:2010aj, 
Nojiri:2010wj} for reviews). Unlike theories 
containing more general  terms quadratic in the curvature, such as 
$R_{ab}R^{ab}$, $R_{abcd} 
R^{abcd}$, or $C_{abcd}C^{abcd}$ (where $R_{ab}, R_{abcd}, C_{abcd}$ are, 
respectively the Ricci, Riemann, and Weyl tensors), $f(R)$ gravity is 
ghost-free \cite{Stelle:1976gc, Sotiriou:2008rp,DeFelice:2010aj, 
Nojiri:2010wj}. 

Continuing, $R^2$ gravity is the high-curvature approximation of 
Starobinsky gravity \cite{Starobinsky:1980te} $f(R)=R+\alpha R^2 $, which 
provided the first (and still preferred) model of inflation in the early 
universe. Pure $R^2$ gravity itself provides a consistent and attractive 
inflationary scenario (\cite{Rinaldi:2014gua}, see also 
\cite{Rinaldi:2014gha, Rinaldi:2015uvu, Tambalo:2016eqr, 
Ferreira:2019zzx, Vicentini:2019etr, Ghoshal:2022qxk, Rinaldi:2023mdf, 
DeAngelis:2024jqc, 
Cecchini:2024xoq}). Moreover, $R^2$ gravity exhibits special symmetries 
(scale invariance and a reduced Weyl symmetry), which are 
preserved if conformal matter with vanishing trace of its stress-energy 
tensor $T_{ab}^\mathrm{(m)}$ is added to the action.

Here we limit ourselves to the gravitational sector of the theory 
and we address the Minkowski limit of $R^2$ gravity, which is known to be 
quite peculiar. Minkowski space is a global solution of $R^2$ gravity, 
yet, the Minkowskian limit of this theory is pathological in various 
respects. First, it has been known since the 1960s 
\cite{Pechlaner:1966dnt} that $R^2$ gravity does not admit a Newtonian 
limit (see also \cite{Stelle:1976gc}).  More precisely, one cannot obtain 
the Newtonian limit by linearizing the fourth order field equations of 
$R^2$ gravity
\be
R\left( R_{ab} -\frac{R}{4} \, g_{ab}\right) 
- \nabla_a \nabla_b R + g_{ab} \Box R 
=   4  \pi T_{ab}^\mathrm{(m)}  \label{fourthorder}
\ee
around Minkowski space, as done in GR \cite{Wald:1984rg}. 
However, it has recently been shown that it is possible to obtain a 
Newtonian limit {\em around de Sitter space} \cite{Nguyen:2023whv}. The 
weak-field  metric  describing a static spherical perturbation of mass $M$ 
on this background turns out to be \cite{Nguyen:2023whv} 
\be 
\d s^2=-\left( 1-\frac{M}{ 3 \Lambda r} -\frac{\Lambda 
r^2}{3} \right)\d t^2 + 
\frac{\d r^2}{  1-\frac{M}{ 3\Lambda r} -\frac{\Lambda 
r^2}{3} } +r^2 \d\Omega_{(2)}^2 \,,\label{Nguyen}
\ee 
where $\d\Omega_{(2)}^2 \equiv \d\vartheta^2+ \sin^2\vartheta \, 
\d\varphi^2$ is the metric on the unit 2-sphere, while  $\Lambda=R_0/4$ is 
the effective cosmological constant of the de Sitter background with 
constant Ricci scalar $R_{0}$. When  one tries to obtain Minkowski space 
as the parameter limit $\Lambda\to 0$, the metric component 
$g_{00}=-1/g_{11}  $  
diverges, making it clear that this is a singular limit. Inspection of  
this line element suggests that the effective gravitational coupling 
$G_\mathrm{eff} \sim 1/\Lambda $, which we recover later in 
Eq.~(\ref{noname}) 
using different arguments. This 
result was anticipated in studies of the weak-field limit of general 
$f(R)$ gravity \cite{Chiba:2006jp, Olmo:2006eh,Kainulainen:2007bt}. 
Consider a spherically 
symmetric, static, non-compact body of mass $M$ embedded in a background 
de Sitter 
universe (usually understood as 
a local approximation to a Friedmann-Lema\^itre-Robertson-Walker (FLRW)  
background), as described by the 
line element 
\begin{eqnarray} \label{weakfieldlineelement}
 \d s^2 & = & -\left[ 1+2\Psi(r) -H_0^2r^2 \right] \d t^2 \nonumber 
\\
&&\nonumber\\
& &+ \left[  1+2\Phi(r) +H_0^2r^2 \right] \d r^2 
 +r^2 \d\Omega^2_{(2)}
\end{eqnarray}
in Schwarzschild coordinates,  where $\Psi(r)$ 
and $\Phi(r)$ are Parametrized Post-Newtonian (PPN) potentials. By 
solving the 
linearized field equations, one obtains
\begin{equation} 
\Psi(r)\simeq  - \frac{4 M}{3 f_0' \, r}  \,, \quad\quad 
\Phi(r)=  \frac{2 M}{3 f_0' \, r}  \,,
\end{equation}
where $f_0' \equiv f'(R_0)$ and $R_0=12 H_0^2$ is the constant Ricci 
scalar of de Sitter space.  The PPN parameter $\gamma$ is, therefore,
\begin{equation}
\gamma =-\frac{\Phi(r)}{\Psi(r)}=\frac{1}{2} \,,
\end{equation}
in gross violation of the experimental bound $\left| 
\gamma -1 \right|< 2.3\cdot 10^{-5} $ 
\cite{Bertotti:2003rm}. However, in high-density environments, the 
chameleon mechanism implicitly contained in many $f(R)$ theories gives  a 
short range to the scalar degree of freedom $f'(R)$, hiding the 
deviations from GR in Solar System and galactic environments. 

The de Sitter background of constant Ricci curvature $R_0$ exists if the 
function $f(R)$  obeys the condition \cite{Barrow:1983rx,Faraoni:2007yn},
\be
R f'(R) -2f(R)=0 \,, \label{trace-vacuum-f(R)}
\ee
which is  satisfied {\em identically} by $R^2$ gravity. Indeed, 
integrating this 
differential equation yields the unique solution $f(R)=R^2$ upon 
imposing $f(0)=0$. This property is not shared  by other $f(R)$ 
theories, in which de Sitter solutions may not exist (e.g., in Starobinsky 
$f(R)=R+\alpha R^2$ gravity), or only exist for special (isolated) 
values 
$R_0$ of the Ricci scalar (this property, called ``$R_0$-degeneracy'' in 
\cite{Casado-Turrion:2023rni, Casado-Turrion:2024esi}, is a consequence 
of scale-invariance).  
Furthermore, the linear stability of de Sitter 
space against gauge-invariant  inhomogeneous perturbations coincides with 
that against homogeneous perturbations \cite{Faraoni:2007yn},
\be
\left[ f'(R) \right]^2 -2f(R) f''(R)  \geq 0 \,;
\ee
this inequality is saturated by $f(R)=R^2$, for all values of the Ricci 
scalar, which means that all de Sitter solutions of $R^2$ gravity 
are (marginally) stable.

Now let us take the limit to Minkowski space of $R^2$ gravity, which 
implies $ f_0' = 2R_0 \to 0$: as a consequence, the PPN potentials $\Psi$ 
and $\Phi$ (which contain $f_0'$ in their denominators) blow up, 
invalidating the weak-field approximation. To the best of our knowledge, 
this feature went unnoticed in the vast literature on $f(R)$ gravity, 
until Ref.~\cite{Nguyen:2023whv} analysed it in detail for pure $R^2$ 
gravity. This problem is caused by the fact that the Minkowski 
background is strongly coupled and gravity is not perturbative around it: 
a strong gravity description is needed. This problem is much more general 
and occurs in {\em all} $f(R)$ gravities in which $f'(0)\to 0$ as $R\to 0$ 
(cf. \cite{Casado-Turrion:2024esi}).

%%%
While, generally speaking, the field content of metric $f(R)$ gravity consists of two massless spin two gravitons (already present in GR) plus the scalar degree of freedom $f'(R)$ \cite{Sotiriou:2008rp,DeFelice:2010aj, Nojiri:2010wj} (claims of the presence of a second scalar breathing mode \cite{Alves:2009eg,Alves:2010ms, RizwanaKausar:2016zgi} have now been refuted \cite{Liang:2017ahj,Moretti:2019yhs}), its behavior around flat spacetime is highly non-trivial. In the Minkowski background, it was historically believed that the two tensor modes do not propagate, leaving only the scalar mode. However, this early conclusion has recently been overturned, showing that the physical spectrum is actually completely empty. The exact derivation of this zero-degrees-of-freedom result has now been confirmed through several independent methods. Determining the number of propagating degrees of freedom on physically relevant backgrounds in a theory of gravity can be subtle and, in the case of $R^2$ gravity, it has been debated for years. 

The issue of the Minkowski limit in pure $R^2$ gravity can be rigorously addressed in terms of propagating physical modes on this background \cite{Alvarez-Gaume:2015rwa, Hell:2023mph, Golovnev:2023zen, Casado-Turrion:2024esi, Karananas:2024hoh, Barker:2025gon, Hell:2025wha, Hell:2025lbl}. In this regard, a systematic analysis has been recently carried out in~\cite{Barker:2025gon}, through a full Hamiltonian approach. As a general rule, $f(R)=R^2$ gravity consistently propagates two (massless) tensor modes and one scalar mode ($N_\text{phy}=3$). However, when the Minkowski background is perturbed, at first glance only the scalar degree of freedom seems to survive, but it ultimately turns out to be non-physical. From the Hamiltonian point of view, this fact is due to the changing nature of the constraints of the model, that go from second to first class \cite{Dirac:1958sq, dirac2013lectures}. Indeed, the ADM decomposition leads to a rewriting of $R$ in the action in terms of the extrinsic curvature $K_{ij}$ containing first time derivatives and a second time derivative term $F_{ij}$. Their promotion to independent velocity fields $K_{ij}\longrightarrow {\cal{K}}_{ij}$ and $F_{ij}\longrightarrow {\cal{F}}_{i j}$ leads to the action $S=S\left[ N,N_i,h_{ij},{\cal K}_{ij},{\cal  F}_{ij},\pi^{i j},\rho^{i j} \right]$ with $N$ the lapse function, $N_i$ the shift vector, and $\pi^{i j}, \rho^{ij}$ Lagrange multipliers (and momenta conjugated to ${\cal K}_{ij}$ and ${\cal F}_{ij}$ respectively). The variation of the action with respect to the Lagrange multipliers $N, N_i, \lambda_{ij}$ yields $10$ primary constraints:
\be
{\cal H} \approx 0\,,  \quad  {\cal H}^i \approx 0\,,  \quad \Phi^{ij} \approx 0\,.
\ee
Here $\lambda_{ij}$ comes from the  traceless nature of ${\cal{F}}_{ij}$ 
and $\Phi^{ij}=\rho^{ij}-h^{ij}\rho/3$.
The Poisson bracket of the smeared version of these quantities results in 
$N_{1\text{st}}=4$ first class constraints and $N_{2\text{nd}}=10$ second 
class constraints due to $\{\Phi^{ij},\Psi^{kl}\}\not\approx 0$, where 
$\Psi^{ij}$ identifies a secondary traceless constraint that guarantees 
the vanishing brackets between the primary constraints (see 
\cite{Barker:2025gon} for details). This allows for the total counting of 
physical degrees of freedom $N_\text{phy}$ through the Dirac-Bergmann 
algorithm:
\be
N_\text{phy}=\frac{1}{2} \left( 
N_{\text{can}}-2N_{1\text{st}}-N_{2\text{nd}} \right)
\ee
with $N_{\text{can}}$ the number of canonical variables. In $f(R)=R^2$ 
gravity, one has $N_{\text{can}}=24$ (given by $h_{ij}$, $\rho^{ij}, \pi^{ij}$, ${\cal K}_{ij}$) which yields $N_\text{phy}=3$ \cite{Barker:2025gon,Buchbinder:1987vp}. 
However, the perturbation around Minkowski space removes the transverse 
part of the momentum constraint and requires 
$\{\Phi^{ij},\Psi^{kl}\}\approx 0$. This results in $N_{1\text{st}}=12$ 
and $N_{2\text{nd}}=0$, reducing to zero the physical degrees of freedom,  
i.e., $N_\text{phy}=0$. Ultimately, this holds at every order of 
perturbation. This is because it actually constitutes a pathological 
feature of every traceless-Ricci (i.e., $R=0$) spacetime, in fact the 
same situation arises also in Schwarzschild and in the 
radiation-dominated FLRW universe. The latter is not a vacuum solution, 
but when sourced by conformal matter $R=0$ then the model falls into this 
category. However, from a dynamical system analysis, the point $R=0$ can 
be crossed by some trajectories of the evolution for spatially flat 
cosmology in pure $R^2$ gravity. This fact suggests once more that the 
perturbation method is inherently pathological (no matter the order) and 
 fails to be predictive. This failure appears also when performing the 
 analysis à la Stueckelberg (\cite{Ruegg:2003ps,Stueckelberg:1938zz}), 
 introducing an additional degree of freedom (in a manifestly covariant 
 way) to exploit a gauge redundancy of the resulting theory 
 \cite{Hell:2023mph}. The full theory propagates the usual 2+1 modes but, 
 on the Minkowski background, the scalar field does not propagate physical 
 modes. Cosmological perturbation theory provides the same discrepancy, 
 again a result of the failure of the perturbative approach, that uses the 
 mass of the scalar field $M$ as perturbation parameter. Indeed, both 
 tensor and scalar modes first order corrections turn out to share the 
 same (strong) coupling scale $L_{\text{str}}\sim 1/M$ that diverges as $M 
\to 0$ \cite{Hell:2023mph}. 

In conclusion, the latter represents a second way in which Minkowski space 
is pathological, i.e., due to strong coupling that comes about when 
considering the field content of the theory and its propagating degrees of 
freedom on a certain background.

A new picture of general scalar-tensor gravity \cite{Faraoni:2021lfc, 
Faraoni:2021jri, Giardino:2022sdv, Faraoni:2022gry, Faraoni:2022jyd, 
Faraoni:2022fxo, Faraoni:2022doe, Miranda:2022wkz, Miranda:2022uyk, 
Giusti:2022tgq, Giardino:2023ygc, Giardino:2023qlu, Faraoni:2023hwu, 
Gallerani:2024gdy, Houle:2024sxs, Karolinski:2024nwp, Faraoni:2025alq, 
Faraoni:2025ufi, Pereira:2025dmk, Bhattacharyya:2025tgp, Faraoni:2025dex, 
Faraoni:2025fjq, Giusti:2026ymb, Pereira:2026xog, Pereira:2026hpi, 
Jarv:2026dbb} (including ``viable'' Horndeski gravity 
\cite{Giusti:2021sku,Nucamendi:2019uen}), which contains $f(R)$ gravity as 
a special case, is currently under development based on an analogy between 
these theories and heat conduction in a relativistic effective dissipative 
fluid. In this analogy, heat dissipation in a relativistic fluid 
corresponds to scalar-tensor gravity approaching GR, while the departure 
from it corresponds to heating due to effective heat sources. In this new 
formalism, an effective ``temperature of gravity'' is defined, to be 
understood as a temperature {\em relative to GR}, which plays the role of 
zero-temperature ``state of equilibrium'' of gravity. This new formalism 
is summarized in Sec.~\ref{sec:2}, while Sec.~\ref{sec:3} reviews pure 
$R^2$ gravity and Sec.~\ref{sec:4} applies the thermal perspective to the 
peculiar Minkowskian limit of this theory.

The thermal formalism, {\em per se}, does not lead to new physics 
for the strongly coupled Minkoswski space of $R^2$ gravity, however there 
are two important motivatiosn for our work. First, the fact that 
strong coupling corresponds to a thermal pathology is needed for  
the new formalism to be trusted in unusual physical situations, which is 
indeed the case. Second, the thermal discussion highlights a physical 
pathology without the need to perform a full Hamiltonian analysis (which 
is, however, needed for a full undeerstanding). Thus, the divergenc eof 
thermal quantites in the new thermal 
formalism can be used as a proxy for singular physics.

\section{The thermal formalism of scalar-tensor gravity} 
\label{sec:2}
\setcounter{equation}{0}

Although the thermal formalism of scalar-tensor gravity has been extended 
to ``viable'' Horndeski theories \cite{Giusti:2021sku, Miranda:2024dhw, 
Nucamendi:2019uen}, here it is sufficient to restrict ourselves to 
``first-generation'' scalar-tensor gravity described by the Jordan frame 
action \cite{Brans:1961sx,Bergmann:1968ve, Nordtvedt:1968qs, 
Wagoner:1970vr, Nordtvedt:1970uv} 
\be
S_\mathrm{ST} = \frac{1}{16\pi}\int d^4 x \sqrt{-g} \left[ \phi R
-\frac{\omega}{\phi} \, \nabla^c \phi \nabla_c \phi -V(\phi) \right] +
S^\mathrm{(m)} \,, \label{action}
\ee
where $\phi>0$ is the Brans-Dicke-like scalar field (approximately the 
inverse of the effective gravitational coupling $G_\mathrm{eff} \simeq 
1/\phi$). $V(\phi)$ is the scalar potential, while the ``Brans–Dicke 
coupling'' $\omega(\phi)$ is required to satisfy $\omega > -3/2$   
to prevent $\phi$ from behaving 
as a phantom field. Finally, $ S^\mathrm{(m)} $ denotes the matter action.
The Jordan frame field equations take the form 
\begin{align}  
 R_{ab}-\frac{1}{2} \, g_{ab}R &= \frac{8\pi}{\phi}  \, 
T_{ab}^\mathrm{(m)} + T_{ab}^{(\phi)} \,, \label{fe1}\\ 
 (2\omega+3) \Box \phi &= 8\pi T^\mathrm{(m)} +\phi \,  V' -2V 
-\frac{d\omega}{d\phi} \, \nabla^c\phi \nabla_c\phi \,,
 \label{fe2} 
\end{align}
with $T_{ab}^{(\phi)}$ the effective stress tensor of the  dissipative 
$\phi$-fluid, 
\begin{eqnarray}
T_{ab}^{(\phi)} \equiv \frac{\omega}{\phi^2} \left( \nabla_a \phi 
\nabla_b 
\phi -\frac{1}{2} \, g_{ab} \nabla^c \phi \nabla_c \phi \right) 
&&\nonumber \\+\frac{1}{\phi} \left( \nabla_a \nabla_b \phi - g_{ab} 
\Box\phi \right) -\frac{V}{2\phi}\,g_{ab} \, , \, \label{fe1.5}
\end{eqnarray}
$T_{ab}^\mathrm{(m)}$ the matter stress–energy tensor
\begin{equation}
\label{set}
T_{ab}^\mathrm{(m)} \equiv-\frac{2}{\sqrt{-g}}\, \frac{\delta
S^\mathrm{(m)} }{\delta g^{ab}} 
\end{equation}
with trace $T^\mathrm{(m)} \equiv T_{ab}^\mathrm{(m)}g^{ab} $. 
The thermal analogy introduced and discussed in Refs.~\cite{Faraoni:2018qdr,
Faraoni:2021lfc,Faraoni:2021jri,Giusti:2021sku,Faraoni:2022gry,
Giardino:2022sdv,Gallerani:2024gdy, Faraoni:2025alq,Faraoni:2025ufi} 
relies on the interpretation of $T_{ab}^{(\phi)}$ as the stress–energy 
tensor of an effective dissipative fluid. It holds when $\nabla^a 
\phi$ is timelike and future-oriented, which is essential  for the 
introduction of  the effective fluid four-velocity
\be
u^a = \frac{ \nabla^a\phi}{\sqrt{ -\nabla_c\phi \nabla^c\phi}}\, .
\ee
Then $T_{ab}^{(\phi)}$ has the form of a dissipative 
fluid stress–energy tensor \cite{Eckart:1940te,Maartens:1996vi, 
Andersson:2006nr}
\be
T_{ab}^{(\phi)} =\rho^{(\phi)} u_a u_b + P^{(\phi)} h_{ab} +q_a^{(\phi)} u_b +q_b^{(\phi)} u_a +\pi_{ab}^{(\phi)} \,,
\ee
where $h_{ab} \equiv g_{ab}+u_a u_b$ is the Riemannian metric on the 
$3-$space experienced by observers comoving with the fluid.
The effective energy density, heat flux density, stress tensor, isotropic 
pressure, anisotropic stresses are, respectively, 
\begin{widetext}
\begin{eqnarray}
 8\pi\,\rho^{(\phi)} &=&  -\frac{\omega}{2\phi^2} \, \nabla^e \phi \nabla_e 
\phi  +  \frac{V}{2\phi} + \frac{1}{\phi} \left( \square \phi -  
\frac{  \nabla^a \phi \nabla^b \phi \nabla_a 
\nabla_b \phi}{ \nabla^e \phi  \nabla_e \phi  } \right)  
\,,\label{effdensity}\\
&&\nonumber\\
 8\pi\,q_a^{(\phi)}   &=& \frac{\nabla^c  \phi \nabla^d \phi}{\phi 
  \left(-\nabla^e \phi \nabla_e \phi \right)^{3/2} } \,  
\Big(  \nabla_d \phi \nabla_c \nabla_a \phi 
- \nabla_a \phi \nabla_c \nabla_d \phi \Big) \,, \label{eq:q}\\
&&\nonumber\\
 8\pi\,\Pi_{ab}^{(\phi)}  &=&  
 \left( -\frac{\omega}{2\phi^2} \, \nabla^c \phi \nabla_c \phi 
-\frac{\Box\phi}{\phi} -\frac{V}{2\phi} \right) h_{ab} +\frac{1}{\phi} \, 
{h_a}^c {h_b}^d \nabla_c \nabla_d \phi \,,  \label{eq:effPi2}\\
&&\nonumber\\
 8\pi\,P^{(\phi)}  & = &  - \frac{\omega}{2\phi^2} \, \nabla^e \phi 
\nabla_e \phi - 
\frac{V}{2\phi} - \frac{1}{3\phi}  \left( 2\square \phi + 
\frac{\nabla^a \phi \nabla^b \phi \nabla_b \nabla_a \phi }{\nabla^e \phi 
\nabla_e  \phi }  \right) \,, \label{effpressure}\\
&&\nonumber\\
 8\pi\,\pi_{ab}^{(\phi)}   &=& \frac{1}{\phi \nabla^e \phi \nabla_e 
\phi } 
\left[  \frac{1}{3} \left( \nabla_a  \phi \nabla_b \phi - g_{ab} 
\nabla^c 
\phi \nabla_c \phi \right) \left(  \square \phi  - 
\frac{  \nabla^c \phi  \nabla^d \phi \nabla_d \nabla_c \phi }{ 
\nabla^e \phi 
\nabla_e \phi }   
\right) \right. \nonumber\\
&&\nonumber\\
&\, & \left. + \nabla^d \phi \left(  \nabla_d \phi \nabla_a \nabla_b 
\phi - 
\nabla_b \phi \nabla_a \nabla_d  \phi - \nabla_a \phi \nabla_d \nabla_b 
\phi +  
\frac{ \nabla_a \phi \nabla_b \phi  \nabla^c \phi \nabla_c 
\nabla_d \phi }{ \nabla^e \phi \nabla_e \phi } \right) \right] \,,
\label{piab-phi}
\end{eqnarray} 
\end{widetext}
where the effective isotropic pressure is the sum of perfect fluid and 
viscous contributions $P^{(\phi)}=P_\mathrm{pf}+P_\mathrm{v}$. The second 
derivatives of $\phi$ give  this effective fluid a dissipative structure. 
Although this decomposition  is general for symmetric rank-two tensors and 
does not by itself carry  physical content \cite{Faraoni:2023hwu}, an 
interesting correspondence  emerges because $ q_a^{(\phi)}$ turns out to 
satisfy Eckart’s constitutive equation for dissipative fluids 
\cite{Eckart:1940te,Maartens:1996vi,Andersson:2006nr} 
\be
q_a = - {\cal K} h_{ab} \left( \nabla^b {\cal T}+ {\cal T} \dot{u}^b
\right) \,,
\ee
where ${\cal K}$ is the thermal conductivity, ${\cal T}$ is the 
temperature, and $\dot{u}^a \equiv u^c \nabla_c u^a$ is the 
fluid four-acceleration (this is Eckart's generalization of the 
Fourier law of heat conduction). A direct calculation 
\cite{Faraoni:2018qdr} shows that 
$q_a^{(\phi)}$ is proportional to $\dot{u}_a$, which allows one to 
identify its coefficient with
\be
{\cal K}{\cal T} = \frac{ \sqrt{-\nabla^c\phi \nabla_c\phi} }{ 8\pi \phi} 
\,, \label{KTdefinition}
\ee
which defines the product of the effective thermal conductivity ${\cal 
K}$ and the 
effective \textit{gravitational temperature} ${\cal T}$. In this thermal 
analogy, 
the scalar-tensor theory corresponds to  an effective dissipative fluid 
with  GR as   
its thermal equilibrium state. Indeed, $\phi=$ const. corresponds to 
${\cal 
K}{\cal T}=0$. When the scalar field is dynamical, we deduce the 
evolution equation for ${\cal KT}$
\begin{align}
\frac{d \left( {\cal K}{\cal T}\right)}{d\tau} = &8\pi \left( {\cal 
K}{\cal T}\right)^2 
-\Theta \, {\cal K}{\cal T} + \frac{ T^\mathrm{(m)} }{\left(  2\omega + 3 
\right) 
\phi} &&\nonumber\\& +\frac{1}{8\pi 
\left( 2\omega + 3 \right)} \left(  V' -\frac{2V}{\phi} 
-\frac{1}{\phi} \, \frac{d \omega}{d\phi} \, 
\nabla^c\phi \nabla_c\phi \right) \label{evolution_general2} \, ,
\end{align}
where $\tau$ is the proper time of the effective fluid and $\Theta\equiv 
\nabla_a u^a$ is its expansion scalar. In this picture, positive 
contributions to the right-hand side of \eqref{evolution_general2} 
\textit{heat} gravity, possibly ending with ${\cal KT}\to \infty$  
(a hot singularity infinitely distant from GR), while negative ones 
\textit{cool} gravity favoring thermal relaxation toward GR, i.e., ${\cal 
KT} \to 0 $  
\cite{Faraoni:2021lfc,Faraoni:2021jri,Giusti:2021sku,Faraoni:2022gry, 
Giardino:2022sdv,Gallerani:2024gdy,Faraoni:2025alq,Faraoni:2025ufi}. 
Ref.~\cite{Faraoni:2025alq} extends this thermal picture to physical 
situations  with $\Box\phi=0$, in which case  the $(\Theta, {\cal K}{\cal 
T})$ plane provides a useful representation of the gravitational 
evolution, turning it into an attractor-repellor thermal mechanism 
towards, 
or from, GR. Indeed, if $\Theta<0$, or if the system starts above the 
critical half-line $8\pi {\cal K}{\cal T}= \Theta>0$, gravity necessarily 
heats up and departs from the GR regime. Conversely, when $\Theta>0$ and 
the system begins below the threshold $8\pi {\cal K}{\cal T}<\Theta$, 
gravity cools and asymptotically approaches GR. This behaviour remains 
valid even when conformal matter with $T^{(m)}=0$ is included, provided 
no potential is present.

\section{$f(R)$ gravity}
\label{sec:3}

For the rest of this manuscript, we focus on $R^2$ gravity which belongs 
to a subclass of scalar-tensor gravity, i.e., metric $f(R)$ gravity 
described by the action 
\be
S= \frac{1}{16\pi } \int d^4 x \sqrt{-g} \, f(R) +S^\mathrm{(m)}
,\label{f(R)action}
\ee
where $f(R)$ is a nonlinear function of the Ricci scalar. Variation with 
respect to the inverse metric $g^{ab}$ yields the fourth order field 
equations \cite{Sotiriou:2008rp,DeFelice:2010aj, Nojiri:2010wj} 
\be
f'(R)R_{ab}-\frac{f(R)}{2} g_{ab} =8\pi T_{ab}^\mathrm{(m)} + \left( 
\nabla_a \nabla_b-g_{ab}\Box\right) f'(R)
\,, \label{metf}
\ee
where primes denote derivatives with respect to $R$.

Taking the trace of Eq.~(\ref{metf}) leads to
\be
\label{metftrace}
\Box f' + \frac{1}{3} \left[ f'(R)R-2f(R) \right] =\frac{8\pi}{3} \, 
T^\mathrm{(m)} \,,
\ee
which shows explicitly that $f'(R)$ propagates as a dynamical scalar 
degree of freedom.

It is well established \cite{Sotiriou:2008rp,DeFelice:2010aj, 
Nojiri:2010wj} that metric $f(R)$ gravity can be rewritten in the form of 
a Brans–Dicke theory with Brans–Dicke parameter $\omega=0$ through the 
identifications 
\be
\phi = f'(R)\, , \qquad V(\phi) = \phi R -f(R) \Bigg|_{ f'(R)=\phi} 
\,.\label{potential}
\ee
In general, this potential cannot be written explicitly in terms of $\phi$.

Physical consistency further imposes  conditions on the function $f(R)$. 
In particular, the requirement $f'(R)>0$  ensures a positive effective 
gravitational coupling and avoids negative-energy  graviton modes. In 
addition, stability under local perturbations demands $f''(R)>0$, 
preventing tachyonic instabilities \cite{Dolgov:2003px,Faraoni:2006sy}.

We now specialize to homogeneous and isotropic  cosmology described by the 
FLRW metric in comoving coordinates $\left( t,r, \vartheta, \varphi \right)$,
\be
\d s^2=-\d t^2 + a^2(t) \left[ \frac{\d r^2}{1-kr^2}+r^2 \left( \d\vartheta^2 +
\sin^2\vartheta \, \d\varphi^2 \right)\right]
\ee
where $k=0,\pm1$ denotes the normalized spatial curvature. In this 
background, the $f(R)$ field equations take the form
\begin{align}
    \label{H_squa}
H^2 =  \frac{1}{3f'}\left[ 8\pi\rho+ \frac{Rf'-f}{2}-3H\dot{R}f''\right] 
-\frac{k}{a^2}  \, ,\\
 2\dot{H} +3H^2  =  -\frac{1}{f'}\Big[ 8\pi P + (\dot{R})^2 f''' 
+2H\dot{R}f'' \nonumber\\+\ddot{R}f''+\frac{1}{2}\left( f-Rf' 
\right)\Big] -\frac{k}{a^2}  \, ,  \label{H_dot} 
\end{align}
where $H \equiv \dot{a}/a$ is the Hubble function and  $\rho$ and $P$ 
denote the energy density and pressure of the 
cosmic matter fluid, respectively.
Within the thermal formulation developed previously, the  evolution of 
${\cal K}{\cal T}$ in vacuum $f(R)$ gravity takes the compact form
\be
\frac{ d\left( {\cal K}{\cal T}\right)}{d\tau} = {\cal K}{\cal T}\left(
8\pi
{\cal K}{\cal T}-\Theta\right) + \frac{ 2f(R)-Rf'(R)}{24 \pi f'(R) } \,.
\label{KTf(R)} 
\ee

In a FLRW universe, the gradient $\nabla^a\phi$ of the Brans-Dicke-like 
scalar field $\phi(t)=f'(R(t))$ is always timelike and the proper time 
$\tau$ of the effective fluid coincides with the comoving time $t$.  
Denoting with an  overdot the differentiation with respect to  $t$, we 
have
\be
\dot{\phi} = f''(R) \dot{R} \,, \label{zzz}
\ee
so that the sign of $\dot{\phi}$  is determined by $\dot{R}$, given the 
stability requirement $f''(R)>0$.

For the fluid interpretation to be consistent, the four-velocity
\be
u^a= \frac{ \nabla^a\phi}{\sqrt{-\nabla^c\phi \nabla_c\phi } } =
\frac{\nabla^aR}{\sqrt{-\nabla^cR \nabla_cR}}
\ee
must be future-directed. In the present coordinates this requires 
$\nabla^0\phi>0$, which translates into the condition $\dot{\phi}<0$ 
 with   $\nabla^0\phi=-\dot{\phi}$. Consequently, according to 
Eq.~\eqref{zzz}, the thermal formalism only applies to FLRW solutions of 
$f(R)$ gravity in which the Ricci scalar
\be
R(t)=6\left( \frac{ \ddot{a} }{a} + \frac{\dot{a}^2 }{ a^2} +\frac{k}{a^2}
\right)
\ee
decreases in  time.

In the FLRW background, we have
\be
{\cal K}{\cal T} = \frac{ |\dot{\phi}|}{ 8\pi \phi} = \frac{ f''(R)
|\dot{R}|}{8\pi f'(R)} \,.
\ee
All solutions with $R=\text{const.}$  correspond to equilibrium states 
with ${\cal K}{\cal T}=0$ \cite{Giardino:2023qlu}.

Before continuing onto the Minkowski limit of $R^2$, it is useful to 
recall the structure of the dynamical phase space in spatially flat FLRW 
cosmology. In scalar–tensor cosmology, the vacuum $k=0$ system can be 
described using the variables $(H,\phi,\dot{\phi})$ \cite{Faraoni:2005vc}. 
$H$ serves as an observable 
phase space coordinate alongside the scalar field $\phi$ and   
$\dot{\phi}$. Although these variables define a three-dimensional space, 
the Hamiltonian constraint restricts physical trajectories to a 
two-dimensional surface embedded in it. In Brans–Dicke form, this 
constraint reads
\be
H^2 = -H \, \frac{ \dot{\phi} }{\phi} + \frac{\omega}{6}  \left(
\frac{\dot{\phi} }{\phi} \right)^2 + \frac{V(\phi)}{6\phi} \,,
\ee
which can be solved for $\dot{\phi}$ as
\be
\dot{\phi} \left( H, \phi \right) = \frac{3\phi}{\omega} \left[ H \pm
\sqrt{ H^2 -\frac{2\omega}{3} \left( \frac{V}{6\phi} - H^2\right) }
\, \right] \,. \label{boh}
\ee
The two branches correspond to distinct sheets of the constraint surface, 
which meet at the boundary of a forbidden region where the square-root 
argument vanishes.

In $f(R)$ gravity the structure simplifies further \cite{deSouza:2007zpn}. As before, the scale factor appears only through $H$ and it is convenient to use $\Theta=3H$ together with $R$ as phase space variables. Since $f'(R)>0$ guarantees invertibility, $R$ is a valid coordinate. The third variable can be chosen as ${\cal K}{\cal T}$ instead of $\dot{R}$, provided one restricts attention to the branch with $\dot{R}<0$ in order to remain consistent with the thermal interpretation.

Because $f(R)$ corresponds to a Brans–Dicke theory with $\omega=0$, the Hamiltonian constraint becomes linear in $\dot{\phi}$, allowing $\dot{R}$ to be eliminated explicitly:
\begin{eqnarray}
\dot{R}\left( \Theta, R \right) &=& \frac{Rf'(R)- f(R)- 6H^2
f'(R)}{6H f''(R)} \nonumber\\
&&\nonumber\\
&=& \frac{Rf'(R)- f(R)- 2\Theta^2 f'(R)/3}{2\Theta f''(R)} \, .\nonumber\\
&&
\end{eqnarray}

As a result, trajectories are confined to the two-dimensional manifold in 
the $(\Theta,R,{\cal K}{\cal T})$ space
\be
{\cal K}{\cal T} \left( \Theta, R \right) = \frac{1}{16 \pi f'(R)} \,
\left| \frac{ Rf'(R)-f(R) -2\Theta^2 f'(R)/3}{\Theta} \right| \,.
\ee
Unlike more general scalar-tensor theories where multiple sheets project 
onto the $(\Theta,R)$ plane, here the projections of trajectories do not 
intersect \cite{deSouza:2007zpn}.

Depending on the functional form of $f(R)$, the physical phase space may 
contain forbidden regions ${\cal F}$ associated with $H^2<0$. In terms of 
$(\Theta,R,{\cal K}{\cal T})$, this condition can be expressed as
\be
Rf'(R)-f(R)+ 16\pi \Theta f'(R) {\cal K}{\cal T} <0 \,.
\ee
Fixed points of the dynamical system correspond to constant curvature 
solutions with $H=H_0$ and $R=R_0=12H_0^2$ 
\cite{Barrow:1983rx,deSouza:2007zpn}, giving
\be
\left( H, R, {\cal K}{\cal T} \right)= \left( \pm \sqrt{ \frac{f_0}{
6f_0'}}, \frac{2f_0}{f_0'}, 0 \right) \,,
\ee
where $f_0 \equiv f(R_0)$ and $f'_0 \equiv f'(R_0)$, which correspond to 
de Sitter configurations, with Minkowski space appearing as the limiting 
case $\left( 0,0,0 \right)$.

Finally, the phase space splits into sectors associated with different 
spatial curvatures. The two-dimensional surface described above separates 
regions corresponding to $k=+1$ and $k=-1$, while the $k=0$ sector forms a 
boundary that cannot be crossed dynamically, reflecting the conservation 
of spatial topology.

\section{Thermal view of the Minkowski limit of $R^2$ gravity} 
\label{sec:4}
\setcounter{equation}{0}

The fact that $R^2$ gravity does not admit a Newtonian limit around 
Minkowski space clearly implies that this theory also lacks  a GR limit 
because GR has  a Newtonian limit, obtained by 
linearizing the Einstein equations around Minkowski space for weak fields 
and slow motions \cite{Wald:1984rg}.  The fact that the two 
massless spin two modes of GR do not propagate in the Minkowski 
background 
reinforces the idea that $R^2$ gravity drastically deviates  from GR at 
the Minkowski limit. In the thermal view of scalar-tensor gravity, this 
would correspond to ${\cal K}{\cal T} $ growing, instead of approaching 
zero as one would naively expect if gravity was going to GR. This is a 
novel point of view on the Minkowskian limit of $R^2$ gravity.

Let us apply the thermal formalism to $R^2$ gravity. 
To preserve scale-invariance, in  $R^2$ gravity it is natural to include 
only conformally invariant matter  with 
$T^\mathrm{(m)}=0$, in which case the evolution equation~(\ref{KTf(R)})  
for ${\cal K}{\cal T}$ simplifies to
\be
\frac{ d\left( {\cal K}{\cal T}\right)}{d\tau} = {\cal K}{\cal T} \left( 
8\pi {\cal K}{\cal T} -\Theta \right) \,,
\ee
which is exactly the situation studied in \cite{Faraoni:2025alq}.

\subsection{$\Lambda\to 0$ limit of de Sitter}

In view of taking the Minkowskian limit, let us consider spatially flat 
FLRW cosmology in $R^2$ gravity, which 
includes de Sitter space as a special case.  The line element is
\be
\d s^2=-\d t^2 +a^2(t) \left( \d x^2 +\d y^2+\d z^2 \right)
\ee
in comoving Cartesian coordinates, where $a(t)$ is the scale factor, $H 
\equiv \dot{a}/a$ is the Hubble function, and an overdot denotes 
differentiation with respect to the cosmic time $t$.

For $R^2 $ gravity, the quadratic potential $V(\phi) = Rf'(R) - f(R) = 
\phi^2/4 $ for the scalar field $\phi=f'(R)$ drops out of 
the field  equation for $\phi$, where it only appears in the combination 
$\phi V'-2V$.  Then, in a FLRW universe, we have
\be
\Box \phi =-\left( \ddot{\phi}+3H \dot{\phi} \right) = 0 \,,
\ee
which admits the well-known first integral
\be
\dot{ \phi} = \frac{C}{a^3} \,,
\ee
where $C<0$ is an integration constant.\footnote{The condition $C<0$ 
ensures $\dot \phi < 0$ (see Sec.~\ref{sec:2}).} Equivalently,   
$\dot{R}=\mbox{const.}/a^3 $. 

de Sitter space corresponds to $\phi=2R=8\Lambda$, where $\Lambda$ is the 
effective cosmological constant defined by the constant value of $R$. Let 
us reason on reaching Minkowski spacetime through the parameter limit 
$\Lambda \to 0$.

The thermal view of $R^2$ gravity in a FLRW background yields
\be 
\K\T=\frac{f''(R) |\dot R|}{8\pi f'(R)}
=\frac{|\dot R|}{8 \pi R} \,, \quad \quad \Theta=3H \, , 
\ee 
where 
\be \dot R=\frac{R 
f'-f-6H^{2}f'}{6Hf''}=\frac{R(R-12H^{2})}{12H} \, . 
\ee 
In de Sitter 
spacetime it is $R = 4\Lambda=12H^{2}$, which implies that in the de 
Sitter limit (excluding the trivial Minkowski subcase) 
\be 
\K \T\to 0 \, , \quad \quad \Theta \to \sqrt{3\Lambda} \,.
\ee 
 The system, therefore, relaxes toward de Sitter 
solutions with $H>0$ that are zero-temperature (GR)  equilibrium states 
of $R^2$ gravity. The $\Lambda \to 0$ limit of de Sitter space  
would be 
an ineffective criterion to describe Minkowski space because  $\K \T=0$ 
identically in de Sitter. Instead, if one 
starts from a FLRW background and performs the Minkowski  limit $R 
\to 0$, $H \to 0$, the expression of the temperature can be simplified 
algebraically before evaluating the limit:
\be
\K\T = \frac{1}{8\pi R} \left| \frac{R(R-12H^{2})}{12H} \right| = 
\frac{|R-12H^{2}|}{96\pi |H|} \, .
\ee
The geometric expression of the Ricci scalar $R = 6\left( \dot{H} + 
2H^2 \right)$ then gives
\be
\K\T = \frac{|\dot{H}|}{16\pi |H|} \, .\label{abu}
\ee
In the Minkowski limit $H \to 0, \dot{H} \to 0$, Eq.~(\ref{abu}) yields 
a structural $0/0$ indeterminate form that cannot be resolved without 
specifying the exact dynamical evolution of $\dot{H}/H$. Thus,  this limit 
is ill-defined, while the expansion $\Theta=3H$ vanishes,
\be
\K\T \longrightarrow \frac{0}{0}, \quad\quad  \Theta \longrightarrow 0 \, .
\ee

\begin{figure*}[t]
    \centering
    \begin{minipage}[t]{0.48\textwidth}
        \centering
        \includegraphics[width=\linewidth]{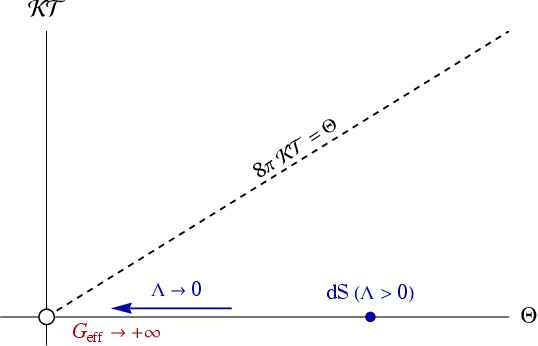}
    \end{minipage}
    \hfill
    \begin{minipage}[t]{0.48\textwidth}
        \centering
        \includegraphics[width=\linewidth]{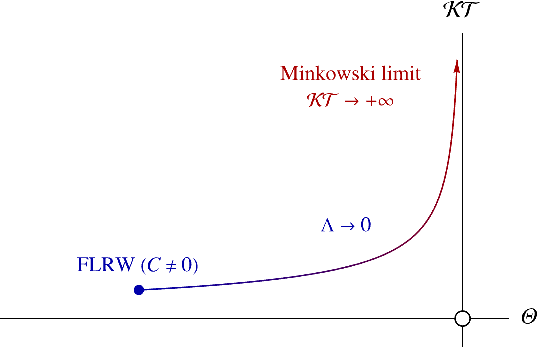}
    \end{minipage}
    \caption{Thermal representation of the Minkowski limit 
for $f(R)=R^{2}$ gravity in the $\Theta-\K\T$ plane. Left panel: in de 
Sitter spacetime a point $(\Theta,\K \T)$ has the form 
$(\sqrt{3\Lambda},0)$    for non-identically vanishing  $\Lambda$, 
therefore belonging to the $\Theta$ axis. The non-dynamical  limit 
$\Lambda\to 0 $ which forces the curvature to switch  off, leads  to 
$(\Theta,\K \T) \to (0,0)$ which, in the thermal formalism  would imply to 
reach GR, but not in this case since here the effective 
gravitational coupling $G_\mathrm{eff} \to +\infty$.\\
Right panel: in FLRW  spacetime $\phi$ changes dynamically and $\K\T$ 
diverges in this limit, (represented by increasing values of $\Theta$) as 
the curvature approaches zero, i.e., $\Lambda \to 0$ (given $C<0$ and 
$a\to 1$). This behavior is shown by  the illustrative (blue and red) 
curve.}
    \label{fig:MLlimit}
\end{figure*}

The effective gravitational coupling strength is
\be
G_\mathrm{eff} = \frac{1}{\phi} = \frac{1}{8\Lambda} \, ,\label{noname}
\ee
which recovers the result of \cite{Nguyen:2023whv} contained in 
Eq.~(\ref{Nguyen}). The 
fact that $G_\mathrm{eff}$ is constant on de Sitter spaces makes it 
clear once again that these ``thermal attractors'' 
are GR points in the $\left( 
\Theta, {\cal K}{\cal T} \right)$ plane. Trying to recover Minkowski 
space and GR from the limit $\Lambda\to 0$ of this parameter is 
delusional:  nothing could be further away from GR than Minkowski space 
with infinitely strong gravity (see Fig.~\ref{fig:MLlimit}).
This issue also emerges  when passing from the Jordan conformal frame 
$\left( g_{ab}, \phi \right)$ to the 
Einstein frame $\left( \tilde{g}_{ab}, \tilde{\phi} \right)$, where the 
conformal transformation $g_{ab} \to \tilde{g}_{ab}=\Omega^2 g_{ab}= \phi 
g_{ab}$ degenerates because the conformal factor $\Omega=\sqrt{\phi}$  
vanishes  at $R=0$. As shown in 
Ref.~\cite{Kehagias:2015ata}, this situation corresponds to the 
``tensionless limit'' in which the effective Planck mass 
$ m_\mathrm{pl}^\mathrm{(eff)} $ vanishes.  This fact reinforces the idea 
that the Minkowski background is associated with an infinitely strongly 
coupled regime which cannot be reached in a perturbative way. 

Strong coupling (in the sense of effective 
field theory) at the Minkowski point has been reported and discussed at 
length  \cite{Alvarez-Gaume:2015rwa, 
Casado-Turrion:2023rni, Casado-Turrion:2024esi, 
Karananas:2024hoh}. Indeed, a more general result holds: strong 
coupling occurs in 
all $f(R)$ backgrounds with constant Ricci scalar $R_0$ which satisfy 
$f(R_0)=0$ and $f'(R_0)=0$; and also in those satisfying $f'(R_0)=0$ 
simultaneously with $f''(R_0)=0$ 
\cite{Casado-Turrion:2024esi}.\footnote{According to a well-known theorem 
by Barrow and Ottewill \cite{Barrow:1983rx}, if $f(0)=0$ and $f'(0)\neq 0$ 
(but $f'(0)>0$ is required to have $G_\mathrm{eff}>0$), the solutions 
reduce to those of GR.}  For $f(R)=R^2$ this conclusion 
agrees with the Hamiltonian  analysis of \cite{Barker:2025gon}. In these 
conditions, the vacuum field equations~(\ref{fourthorder}) reduce to 
\begin{align}
    f'\left( R_0 \right) \, R^{(0)}_{ab} = \frac{ f\left( R_0 \right)}{2} 
\, g^{(0)}_{ab}  \,,
\end{align} 
and the trace equation (\ref{trace-vacuum-f(R)}) degenerates. Perturbing 
the background, we have $g_{ab}= g_{ab}^{(0)} + h_{ab}$ and 
\begin{align}
     f''\left( R_0 \right) \left[ R^{(0)}_{ab} - \left( \nabla_a 
\nabla_b  - g^{(0)}_{ab} \Box \right) \right] R^{(h)} + \mathcal{O}(h^2)=0 
\,,
\end{align}
with trace 
\begin{align}
     f'' \left( R_0 \right) \left( \Box + \frac{R_0}{3} \right)  
R^{(h)} + \mathcal{O}(h^2)=0 \,.
\end{align} 
Now, when $f(R_0)=0$ and $f'(R_0)=0$, the scalar field 
potential~(\ref{potential})   
evaluated on the background vanishes. The situation mimics that of $R^2$ 
gravity already discussed.  

In the context of 
the thermal formalism, situations have been pointed out in which the 
divergence of  $G_\mathrm{eff} $ 
leads to ${\cal K}{\cal T}\to +\infty$, with the lesson that 
singularities of the effective gravitational coupling are to 
be treated similarly to spacetime singularities  
\cite{Giardino:2023qlu}. This conclusion rephrases the 
strong coupling problem of effective field theory. 
It turns out that Minkowski spacetime is strongly coupled in all $f(R)$ 
gravities for which  $f(0) = 0 $ and $f' (0) = 0$ 
\cite{Casado-Turrion:2023rni}. By contrast, no pathologies emerge if both 
$f'(R_0)>0$ and $f''(R_0)>0$.    
We now have  for fixed $C<0$ and $\phi\to 0^{+}$, 
\be
{\cal K}{\cal T} = \frac{ |\dot{R}|}{8\pi R} = 
\frac{|C|}{8 \pi a^3 \phi} 
\to +\infty
\ee
in the Minkowski limit $a\to 1$, $\Lambda \to 0$, and $\phi\to 0$.  In 
other words, $R^2$ gravity does not converge to GR but deviates 
infinitely far from it in this limit, and the divergence is due to the 
vanishing of $\phi$, corresponding to the strong coupling $G_\mathrm{eff} 
\to +\infty$ (this is an exact, not a perturbative argument). To 
visualize the situation, 
consider the $\left( \Theta, {\cal K}{\cal T}\right)$ plane 
(Fig.~\ref{fig:MLlimit}), which was 
shown in \cite{Faraoni:2025alq} to be very useful for the study of 
electrovacuum Brans-Dicke gravity, not only in FLRW cosmology, but in 
general spacetimes. 

The $\Theta>0 $ half-axis is composed of de Sitter 
states at ${\cal K}{\cal T}=0$, corresponding to all possible values of 
$\Lambda>0$. Attempting to reach the Minkowski origin $\left(\Theta, {\cal 
K}{\cal T} \right)=\left( 0,0 \right)$ by taking the limit $\Lambda\to0$ 
and moving to the left along the $\Theta$-axis (which is a parameter 
limit and does not describe dynamical evolution in time) does 
actually lead to the   origin. Indeed, there exists a family 
of solutions satisfying 
\be
\left( \Theta,\K \T \right) = \left( 
\sqrt{3\Lambda},0 \right) \xrightarrow{\Lambda \to 0^{+}}(0,0) \,.
\ee
However, this point is pathological because it exhibits a diverging 
coupling $G_\mathrm{eff} $ (i.e., $\phi\to 0$), which prohibits to 
interpret that point as a   
regular GR equilibrium state with finite gravitational 
coupling, and it should be excluded by the domain of the regular thermal 
quantities, as done in Fig.~\ref{fig:MLlimit} (left panel). By contrast, 
in the non-stationary FLRW scenario (Fig.~\ref{fig:MLlimit},  
right panel) ${\cal K}{\cal T}$ never reaches the vertical 
axis, but diverges to $+\infty$, i.e., 
infinitely far from GR, {\color{red}at $\Theta=0$} as 
$\Theta\to 0^{-}$.  This 
divergent behavior, which moves the system completely away from GR, is 
consistent with the lack of propagating tensor degrees of freedom in the 
Minkowski background of $R^2$ gravity, and the absence of a Newtonian 
(therefore, of a GR) limit \cite{Pechlaner:1966dnt,Nguyen:2023whv} at the 
Minkowski point.
 
\subsection{Phase space}

For FLRW universes in $R^2$ gravity, we have  
\begin{align}
    &\dot{\Theta}=3\dot{H}\,,\\
    &\dot{R}=\frac{R^2}{4\Theta}-\frac{\Theta R}{3} \,,\\
    &\mathcal{KT}\left( \Theta, R \right )=\frac{1}{32\pi}\Big| 
\frac{R}{\Theta}-\frac{4\Theta}{3}\Big| \,,
\end{align}
where $ \left( \mathcal{KT}, \Theta, R \right)$ is the 3-dimensional phase 
space. GR corresponds to ${\cal K}{\cal T}=0$ and to  
$R=4 \Theta^2/3 $. The Minkowski limit  is not straightforward also from 
the point of view of the phase space of FLRW cosmology. 
Ref.~\cite{Barker:2025gon} studies  
the dynamical system for  spatially flat FLRW cosmologies and  concludes  
that the point $R=0$ can be  dynamically 
crossed by the orbits of certain solutions  when 
$\dot{R}>0$ and $H>0$, or when $\dot{R}<0$ and $H<0$ (which is the time 
reversal of the previous situation). The latter case is 
of interest to us because  our thermal formalism applies, and we are 
then forced to consider contracting universes in  
the region $\Theta<0$ where ${\cal K}{\cal T}$ can only increase and 
gravity depart from GR.   

Let $ R( t_0)=0 $,  where $t_0$ is the comoving time when the 
point $R=0$ is crossed. Linearizing near  this crossing time, 
\begin{align}
    R(t)\simeq \dot{R}\left( t- t_0 \right) \,,
\end{align}
and Ref.~\cite{Barker:2025gon} establishes that a crossing is possible 
only if 
$\dot{R}(t_0) \neq 0 $. In the thermal  
formalism, we then have
\begin{align}
    \mathcal{KT}\sim\frac{ 
|\dot{R}(t_0)|}{8\pi| t -t_0|| \dot{R}|} = 
\frac{1}{8\pi| t- t_0|}
\end{align}
and 
\begin{align}
\lim_{t\rightarrow t_0}\mathcal{KT}=+\infty \,.
\end{align}
Therefore, a trajectory crossing $R=0$ appears in the $\left( \Theta, 
{\cal K}{\cal T}\right)$ plane (which is not a phase space) as a 
trajectory reaching  infinite $\mathcal{KT}$. 
Since the orbit in phase 
space is 
continuous and enters the Minkowski point (where $H=0$, $R=0$) from $H<0$ 
and $\dot{R}<0$,  it 
must re-emerge with $H>0$ and, using time-reversal, $\dot{R}>0$ according 
to \cite{Barker:2025gon}; the trajectory would have to re-emerge 
in this region, where the thermal formalism cannot be applied. 

For the crossing trajectories considered here, the signs of
$H$ and $\dot R$ are preserved across $R=0$
\cite{Barker:2025gon}. On the branch with $\dot R<0$,
the continuation enters $R<0$, hence $\phi=2R<0$,
and therefore leaves the positive-coupling domain
in which the thermal formalism is defined.

The thermal formulation distinguishes the strength of the
gravitational coupling from its fractional rate of change.
Indeed, in FLRW spacetime,
\be
\K\T=\frac{1}{8\pi}
\left|\frac{\dot G_{\mathrm{eff}}}{G_{\mathrm{eff}}}\right|\,.
\ee
The exact de Sitter family and the non-stationary FLRW
trajectories considered here both lead to a divergent
effective coupling, but exhibit different thermal behavior.
Consequently, a vanishing thermal quantity along a family
of backgrounds is not sufficient to establish a regular GR
limit: the gravitational coupling must also remain finite
and positive. For regular crossings with
$\dot R(t_0)\neq0$, the asymptotic scaling
$\K\T\sim[8\pi(t_0-t)]^{-1}$ additionally quantifies
the approach to the boundary from the $\phi>0$ side
in terms of the remaining cosmic time.
The contribution of the thermal description is therefore
an exact characterization of these limiting procedures
and of the conditions required to interpret thermal
equilibrium as a regular GR background.

\subsection{Cosmological perturbations}

Finally, the drastic departure of $R^2$ gravity from GR at the Minkowskian 
limit manifests itself also in the divergence of thermal quantities for 
cosmological perturbations. The thermal view of cosmological perturbations 
around a spatially flat FLRW universe in first-generation (Jordan frame) 
scalar-tensor gravity was studied in detail in 
Ref.~\cite{Pereira:2026xog}, demonstrating explicitly an Eckart 
dissipative fluid framework for these perturbations. We refer to 
~\cite{Pereira:2026xog}, adapting the relevant formulae to the specific 
case of $R^2 $ gravity.

The metric and scalar field perturbations $A,\psi, \chi_{ij}, \varphi $ 
for a spatially flat FLRW  universe in the Newtonian gauge are given by 
\be
ds^2 = -\left(1+2A \right) dt^2 + a^2\left[ \left( 1-2\psi\right) 
\delta_{ij}  +\chi_{ij} \right] dx^i dx^j \,,
\ee
\be
\phi = \bar{\phi}(t) +\varphi \left(t, \vec{x} \right) \,,
\ee
where $i, j=1,2,3$, $A$ and $\psi$ are scalar potentials, and the 
$\chi_{ij}$ (satisfying 
$\partial^i \chi_{ij}=0$, ${\chi^i}_i=0$) describe 
gravitational waves, while  $\dot{ \bar{\phi}}<0$ to ensure that $\nabla^a 
\bar{\phi}$ is future-oriented. 

The first order perturbation of ${\cal K}{\cal T}$ is found to be 
\cite{Pereira:2026xog} 
\be 
\delta \left( {\cal K}{\cal T} \right) = \frac{1}{8\pi \bar{\phi} } \left( 
\dot{ \bar{\phi} } A- \dot{\varphi}  
+ \frac{ \dot{\bar{\phi}}   }{\bar{\phi} } \, \varphi \right) \,. 
\label{deltaKT}
\ee
Although $R^2$ gravity and its Minkowskian limit were not contemplated in 
\cite{Pereira:2026xog}, it is now easy to consider them. $R^2$ gravity  is 
equivalent to a 
Brans-Dicke theory with $\phi=f'(R)=2R$, 
$\omega=0$, and potential $ V(\phi)=Rf'(R) - f(R)\bigg|_{\phi=f'(R)}  =  
\phi^2/4 $. As already seen, the Minkowskian limit corresponds to the 
strong coupling limit $ \bar{\phi} \to 0$, in which not only $ {\cal 
K}{\cal T}  \to +\infty$, but also 
$\delta \left( {\cal K}{\cal 
T} \right) $ diverges (cf.  Eq.~(\ref{deltaKT})) and the 
perturbative approximation breaks down.\\

\section{Conclusions}
\label{sec:5}
\setcounter{equation}{0}

A novel point of view on the peculiar Minkowskian limit of $R^2$ gravity 
comes from the new thermal picture of scalar-tensor gravity. In this 
formalism, the anomalies of $R^2$ gravity at the Minkowski limit (absence 
of a Newtonian limit, therefore of a GR limit, loss of the two GR 
propagating degrees of freedom, divergence of PPN potentials, and 
divergence of cosmological perturbations) acquire a 
new meaning. These apparently pathological phenomena occur, and even make 
sense, because in this limit $R^2$ gravity diverges infinitely far away 
from GR in the thermal description, instead of converging to it as naively 
expected. This behaviour 
is a result of the fact that $R^2$ gravity becomes strongly coupled in the 
Minkowski background, as described by the fact that ${\cal K}{\cal 
T}\to +\infty$  along the non-stationary FLRW paths  
considered here. As anticipated in \cite{Giardino:2023qlu}, 
singularities of the effective gravitational coupling should be treated 
similarly to spacetime singularities, which rephrases the strong coupling 
problem of field theory in thermal terms. This argument is 
non-perturbative. Thus, the thermal view of scalar-tensor  gravity 
contributes to resolving a puzzle of $R^2$ gravity and of  all $f(R)$ 
theories on solutions with $R_0=0$, $f(R_0)=0$, and $f'(R_0)=0$.
In the Hamiltonian formulation of $f(R)$ gravity, the canonical 
constraint structure degenerates at surfaces where $f'=0$ and $f''=0$, and 
the $R=0$ solution of $R^2$ gravity is a special case of the wider 
occurrrence of a phase space singularity in $f(R)$ gravity 
\cite{Glavan:2026dfv}.  In most $f(R)$ theories, these singular surfaces 
are not solutions and dynamically crossing them is problematic to say the 
least. It may well be that the blow-up of ${\cal K}{\cal T}$ at these 
surfaces is the thermal counterpart of the Hamiltonian rank change 
\cite{Glavan:2026dfv}: this question for more general $f(R)$ theories will 
be investigated in the future.

Although the thermal formalism does not predict new physics with 
respect to what it is already known for the Minkowski solution of $R^2$ 
gravity, the association of strong coupling with a thermal pathology may 
be useful in the future to detect similar situations in more complicated 
scalar-tensor gravities. Finally (and perhaps more important from the 
point of view of its developers), the divergence of thermal quantities at 
strong coupling renews the confidence in this formalism, which seems to 
adequately signal physical pathologies.

\begin{acknowledgments} 

We are grateful to David M. J. Vokrouhlick{\'y} for a discussion 
and for pointing out typographical errors in a previous version of this 
work. The work of L.G. has been carried out in the framework of 
activities of the National Group of Mathematical Physics (GNFM, INdAM). 
V.F. is partially supported by the Natural Sciences \& Engineering 
Research Council of Canada (Grant No. 2023-03234).

 \end{acknowledgments}

%\begin{appendices}
%\section{Proof of Eq.~(\ref{ToProve}) }
%\label{Appendix:A} 
%\renewcommand{\theequation}{A.\arabic{equation}} 
%\setcounter{equation}{0}
%\newline
%\end{appendices}

\end{document}